\documentclass{aa} 

\usepackage{
graphicx,
txfonts,
color,
natbib,
lscape
}

\bibpunct{(}{)}{;}{a}{}{,} 	

\begin{document}

\title{Ultraviolet and visible photometry of asteroid (21) Lutetia using the
Hubble Space Telescope
\thanks{ÒSupport for this work was provided by NASA through
a grant from the Space Telescope Science Institute (program \#11957),
 which is operated by the Association of Universities for Research in Astronomy, 
 Incorporated, under NASA contract NAS5-26555.}
}
\author{
H. A. Weaver\inst{1}
\and P. D. Feldman\inst{2}
\and W. J. Merline\inst{3}
\and M. J. Mutchler\inst{4}
\and M. F. A'Hearn\inst{5}
\and J-. L. Bertaux\inst{6}
\and L. M. Feaga\inst{5}
\and J. W. Parker\inst{3}
\and D. C. Slater\inst{7}
\and A. J. Steffl\inst{3}
\and C. R. Chapman\inst{3}
\and J. D. Drummond\inst{8}
\and S. A. Stern\inst{9}
}
\institute{
Johns Hopkins University Applied Physics Laboratory,
Space Department,
11100 Johns Hopkins Road,
Laurel, MD 20723-6099 USA
\email{hal.weaver@jhuapl.edu}
\and
Johns Hopkins University,
Department of Physics and Astronomy,
3100 N. Charles Street,
Baltimore, MD 21218 USA
\and Southwest Research Institute, 
Department of Space Studies, Suite 300, 
1050 Walnut Street, 
Boulder CO 80302-5150 USA
\and Space Telescope Science Institute,
3900 San Martin Drive,
Baltimore, MD 21218 USA
\and Department of Astronomy, 
University of Maryland, 
College Park MD 20742-2421 USA
\and Service dÕAeronomie, 
R\'{e}duit de Verri\`{e}res -- BP3, 
Routes des Gatines, 
91371 Verri\`{e}res le Buisson C\'{e}dex, France
\and Southwest Research Institute, 
P. O. Drawer 28510, 
San Antonio TX 78228-0510 USA
\and Starfire Optical Range,
Air Force Research Laboratory,
Kirtland Air Force Base,
Albuquerque, NM 87117 USA
\and Southwest Research Institute,
Space Division,
1050 Walnut Street, 
Boulder, CO 80302-5150 USA
}

\date{Received 23 December 2009 / Accepted 27 June 2010}

\abstract
{The asteroid (21) Lutetia is the target of a planned close encounter by the \emph{Rosetta} spacecraft in July 2010. To prepare for that flyby, Lutetia has been extensively observed by a
variety of astronomical facilities.}
{We used the \emph{Hubble Space Telescope (HST)} to determine the albedo of Lutetia over a wide wavelength range, extending from $\sim$1500~\AA\ to
$\sim$7000~\AA.} 
{Using data from a variety of \emph{HST} filters and a ground-based visible light spectrum,
we employed synthetic photometry techniques to derive absolute fluxes for
Lutetia. New results from ground-based measurements of Lutetia's size and shape were used to convert the absolute fluxes into albedos.} 
{We present our best model for the spectral energy distribution of Lutetia over the
wavelength range 1200--8000~\AA. There appears to
be a steep drop in the albedo (by a factor of $\sim$2) for wavelengths
shorter than $\sim$3000~\AA. Nevertheless, the far ultraviolet albedo of Lutetia ($\sim$10\%)
is considerably larger than that of typical C-chondrite material ($\sim$4\%).
The geometric albedo at 5500~\AA\ is 16.5 $\pm$ 1\%.
} 
{Lutetia's reflectivity is not consistent with a metal-dominated surface at infrared or radar
wavelengths, and its albedo at all wavelengths (UV-visibile-IR-radar)
is larger than observed for typical primitive, chondritic material.
We derive a relatively high FUV albedo of $\sim$10\%, a result that will
be tested by observations with the \emph{Alice} spectrograph during the
\emph{Rosetta} flyby of Lutetia in July 2010.}

\keywords{Minor planets, asteroids}

\titlerunning{{\it HST} UV-visible photometry of Lutetia}
\authorrunning{Weaver et al.}

\maketitle

\section{Introduction}

The {\it Rosetta} spacecraft was launched on 2~March~2004 and is heading toward an historic encounter with comet \mbox{67P/Cheryumov-Gerasimenko} (67P/C-G) in 2014. Along the way, {\it Rosetta} has flyby encounters with two asteroids: the spacecraft passed at a distance
of 803~km from (2867)~Steins on 5~September~2008, and an encounter with (21)~Lutetia is currently being planned for a flyby distance of $\sim$3000~km on 10~July~2010.
\citet{Barucci:2007} provide an excellent summary of the pre-2006 data accumulated
for both of these asteroids.
The first results from the {\it Rosetta} flyby of Steins have recently been published
in {\it Science} \citep{Keller:2010} and in a special volume of 
{\it Planetary and Space Science} (2010).

To assist with the planning of the Lutetia flyby, we made observations of this asteroid
with the {\it Hubble Space Telescope (HST)} in late-November and early-December 2008,
near the time of opposition when both the solar phase angle and the geocentric distance
were minimized. The objectives of the program were:

\begin{enumerate}

\item To measure the far ultraviolet \mbox{($\lambda \approx 1500$ \AA)} albedo of Lutetia
to enable better planning of the flyby observations by the
{\it Alice} ultraviolet (UV) spectral imager \citep{Stern:ssr2007}.

\item To measure the near UV 
\mbox{($\lambda \approx 2000-3000$ \AA)} and visible 
\mbox{($\lambda \approx 4000-7000$ \AA)}
albedo, thereby providing the spectral energy distribution across a wide
wavelength range and possibly yielding improved insights into the nature of 
the surface composition and taxonomic class of Lutetia.

\item To search for dust debris and companions near Lutetia that might pose
a hazard to the {\it Rosetta} spacecraft, using deep, visible-band observations.
 
\item To use spatially resolved {\it HST} images of Lutetia to constrain its size and shape.
 
 \end{enumerate}
 
In this paper, we focus on the first two objectives, providing the best available pre-flyby
estimates of the UV-to-visible spectral energy distribution of Lutetia.
The other objectives will be discussed in separate, future publications.

\section{Observations, Data Reduction, and Results}

We were granted Director's Discretionary Time (program ID 11957) on {\it HST} to perform filter
photometry of Lutetia in support of the {\it Rosetta} flyby
in July 2010.
We were initially allocated 5 orbits (synonymous with ``visits'' in this case) of observing
time, which were executed successfully on 30~November 2008, when Lutetia's heliocentric
distance ($r$) was 2.42~AU, the geocentric distance ($\Delta$) was 1.44~AU, and 
the solar phase angle ($\phi$) was 0\fdg47--0\fdg48.
When a preliminary analysis of those data suggested the presence of a previously
unknown companion to Lutetia (subsequently determined to be an optical ghost),
we were allocated 2 more orbits of {\it HST} time, one of which executed successfully
on 15~December 2008 ($r = 2.45$~AU, $\Delta = 1.50$~AU, $\phi = 7\fdg52$)
and the other executed successfully one day later
($r = 2.45$~AU, $\Delta = 1.50$~AU, $\phi = 7\fdg98$).
Table~\ref{table:log} provides a log of all the {\it HST} observations, detailing the
rootnames for the data files, the time span of each visit, the instrument
used, the filters employed, the image exposure times, and the objectives
of each visit.

For our program, we employed two different instruments: the Planetary Camera (PC) mode
of the Wide Field Planetary Camera~2 (WFPC2) was used for the filter photometry
covering the near-ultraviolet (NUV) and visible wavelength ranges, while the
Solar Blind Channel (SBC) of the Advanced Camera for Surveys (ACS) was
used to measure the far-ultraviolet (FUV) flux.
The WFPC2/PC has square pixels 0\farcs046 on a side, and the
ACS/SBC has (after {\it drizzling} to remove geometrical distortion) 
square pixels 0\farcs025 on a side.
In both cases, we used the standard calibrated images from the processing pipeline
operated by the Space Telescope Science Institute (STScI).

\begin{figure*}
\begin{minipage}[ht]{.5\textwidth}
	\begin{center}
	\includegraphics[angle = 0,width = \textwidth]
	{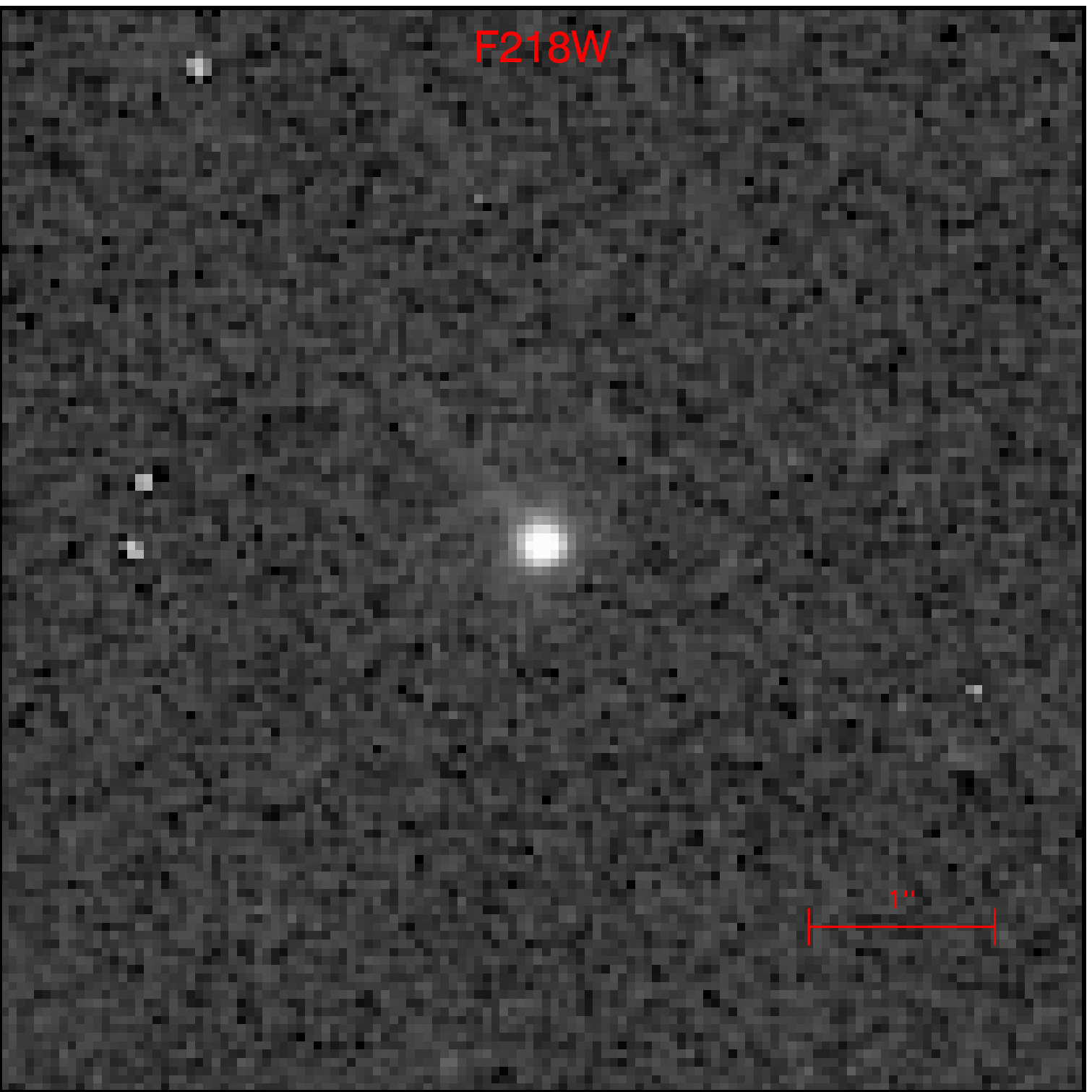}
	\end{center}
\end{minipage}
\hfill
\begin{minipage}[ht]{.5\textwidth}
	\begin{center}
	\includegraphics[angle = 0,width = \textwidth]
	{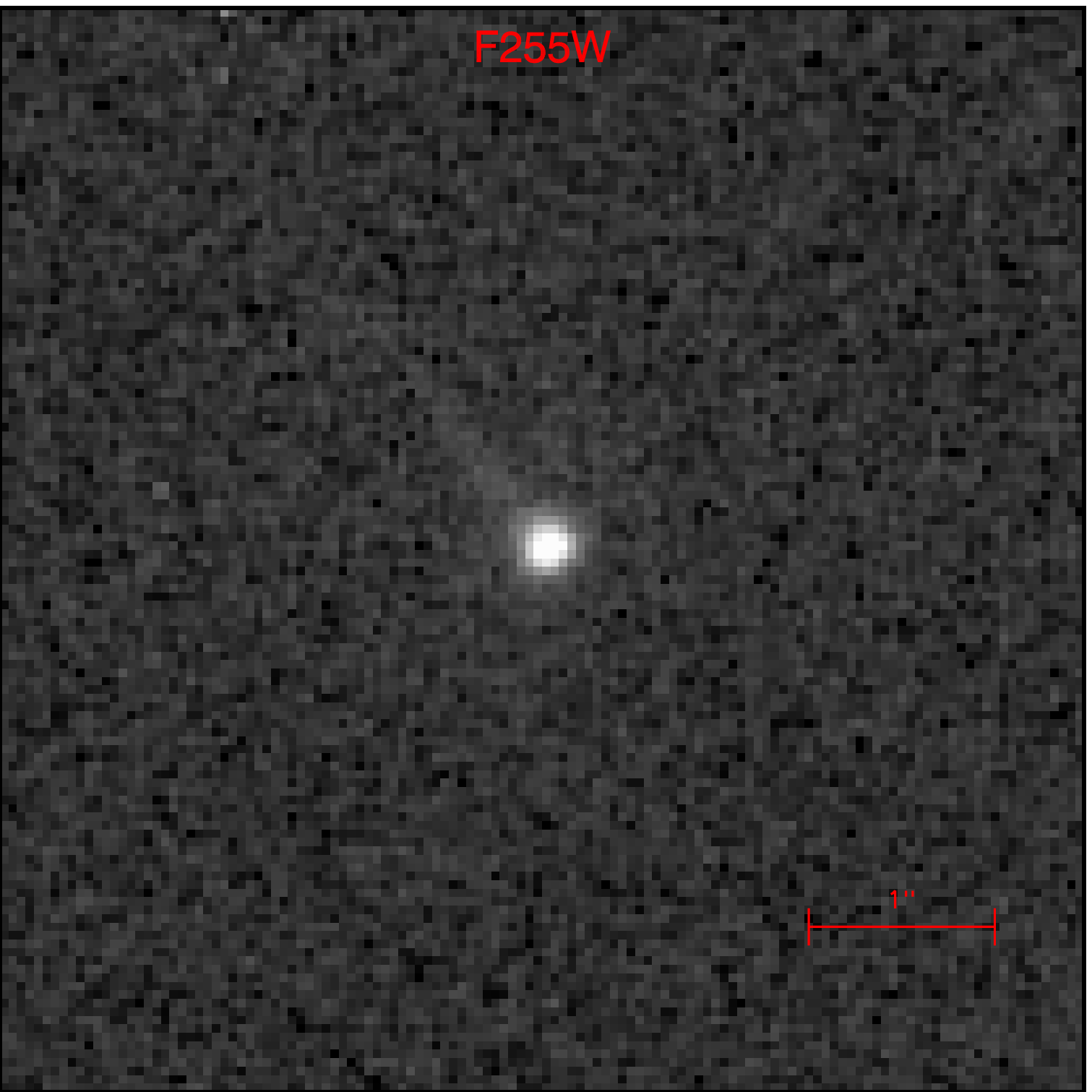}
	\end{center}
\end{minipage}
\begin{minipage}[ht]{.5\textwidth}
	\begin{center}
	\includegraphics[angle = 0,width = \textwidth]
	{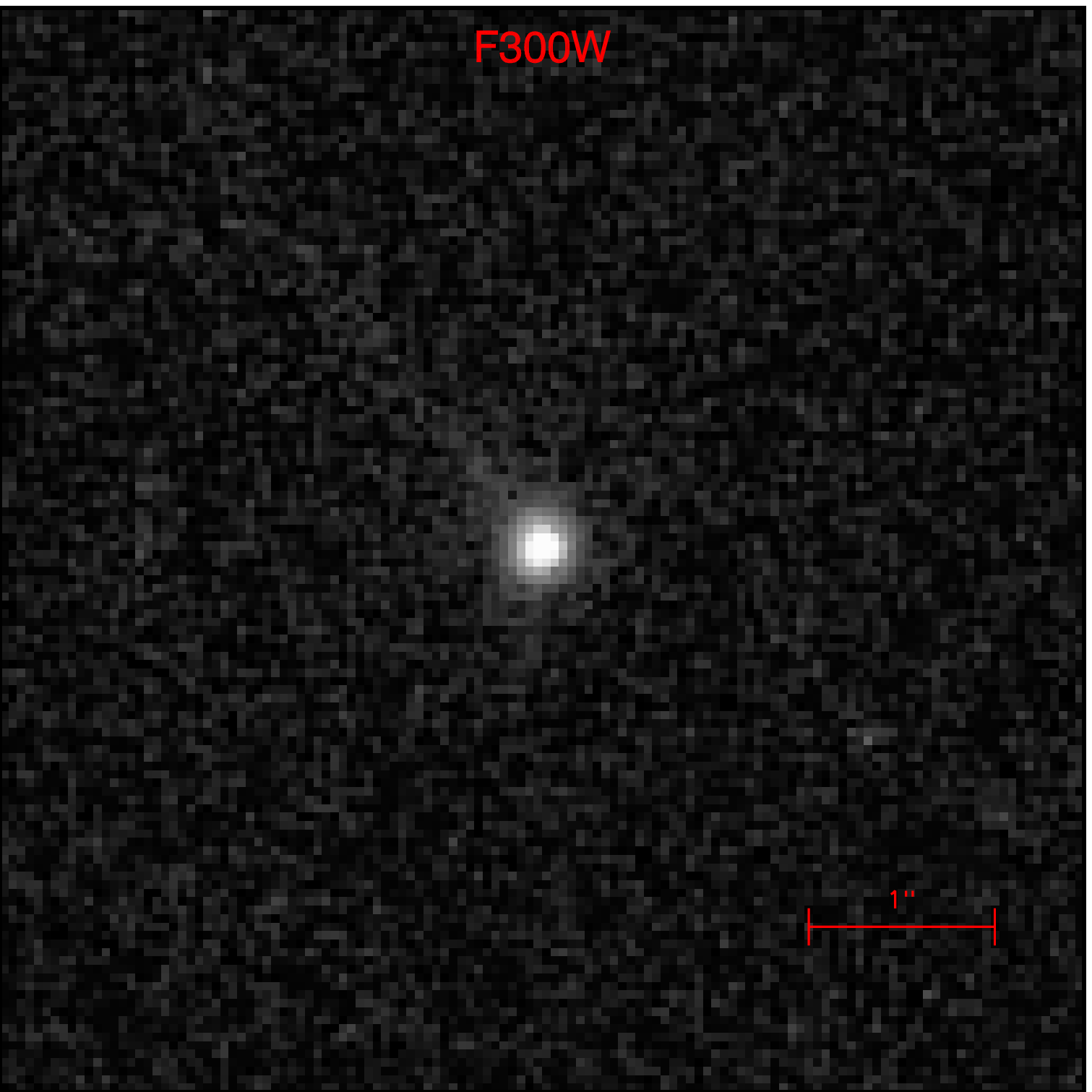}
	\end{center}
\end{minipage}
\hfill
\begin{minipage}[ht]{.5\textwidth}
	\begin{center}
	\includegraphics[angle = 0,width = \textwidth]
	{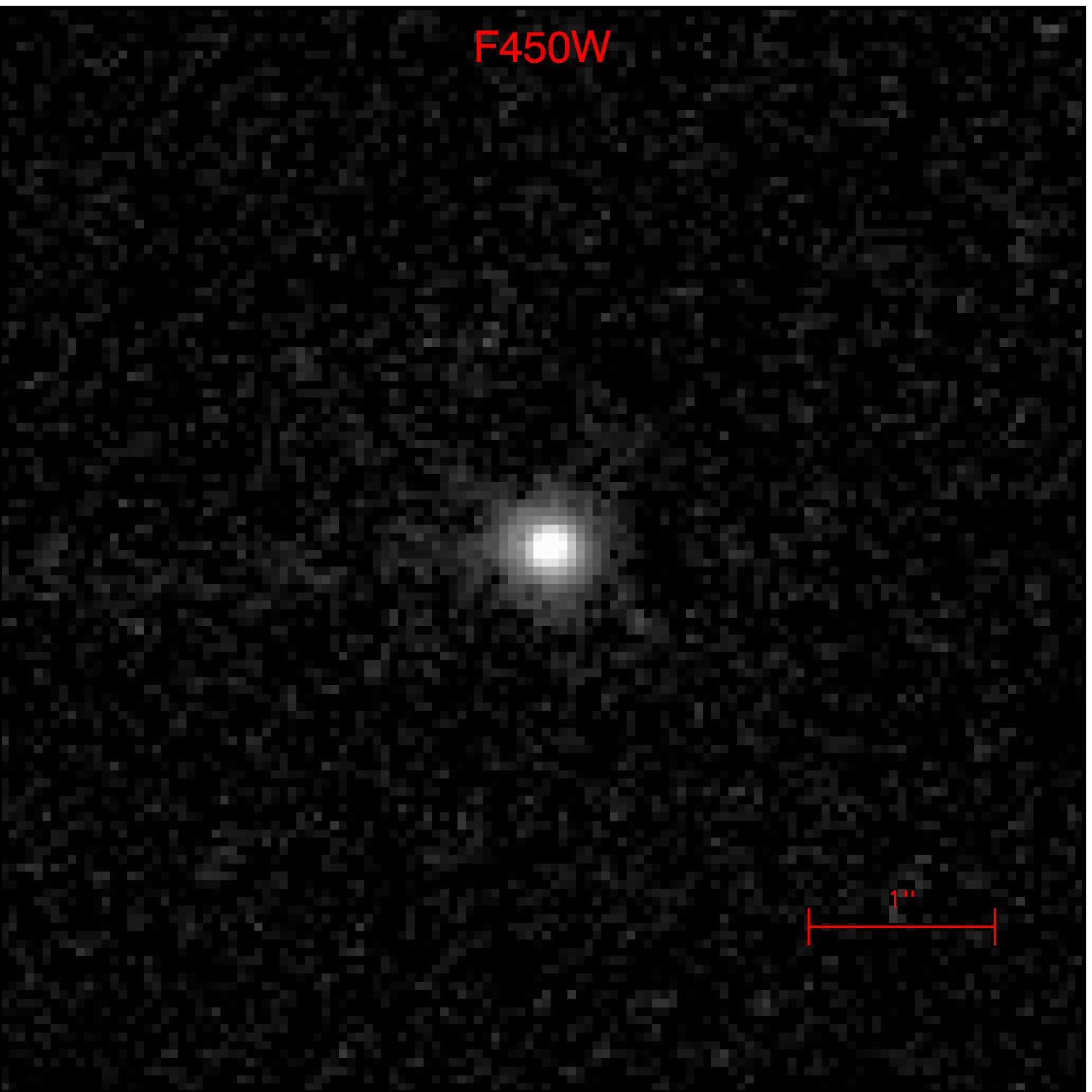}
	\end{center}
\end{minipage}
\caption{{\it HST/WFPC2} images of asteroid 21 Lutetia taken through the F218W,
F255W, F300W, and F450W filters (see identifying labels on the images). 
Each image is a \mbox{128 $\times$ 128} pixel subsection (each image is 5\farcs89 across)
of the full image and is displayed using an asinh
intensity stretch (similar to logarithmic) ranging from approximately zero to the 
maximum intensity in each image.
All images have been rotated so that celestial north points up and east is to the left.
The scale bar is 1\arcsec\ across, which subtends 1088~km at Lutetia on December~15
(when the F450W image was taken), and 1044~km on November~30 (when all the other images
were taken). Each image is a composite of at least two separate images taken with
Lutetia centered at two different locations on the CCD.
The ``tails'' apparent on two of the images are low-level artifacts caused by
degraded charge transfer efficiency in the {\it WFPC2} CCD.}
\label{fig:wfpc2}
\end{figure*}

\begin{figure*}[ht]
\resizebox{\hsize}{!}{
\includegraphics[width=\textwidth]
{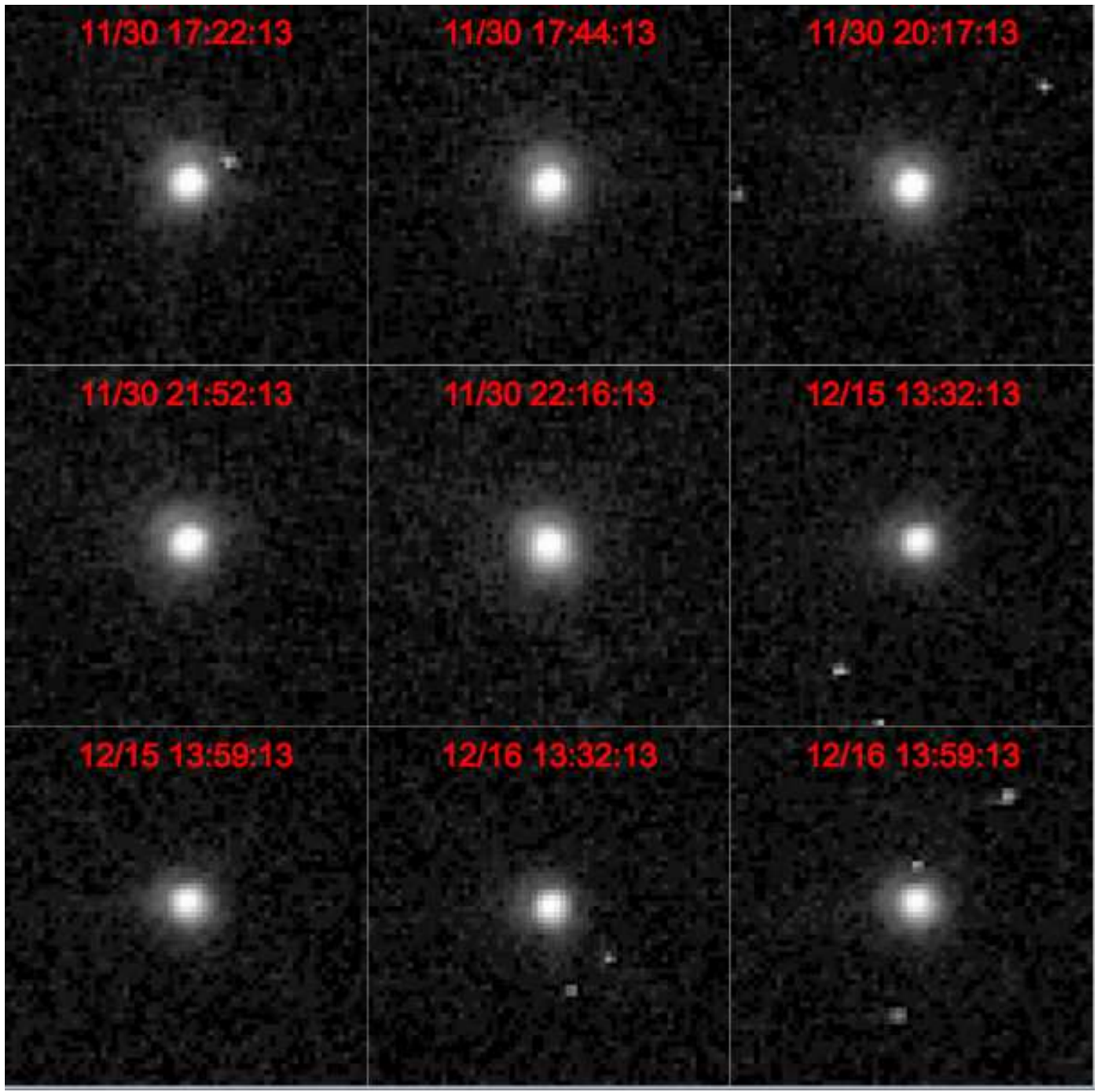}
}
\caption{
{\it HST/WFPC2} images of Lutetia taken through the F606W filter with an exposure
time of 0.11~s and rotated so that celestial north points up and east is to the left.
Each image is a \mbox{64 $\times$ 64} pixel subsection (each image is 2\farcs94 across)
of the full image and is displayed using an asinh
intensity stretch (similar to logarithmic) ranging from approximately zero to the 
maximum intensity in each image.
Cosmic ray events are evident in some of the images as clusters of
bright pixels.
The image start times (UT) are displayed in each frame.
}
\label{fig:f606w}
\end{figure*}

Because multiple images were obtained for the WFPC2 observations, at two different positions
on the detector (to mitigate the effects of cosmic rays and bad CCD pixels), 
we generally combined them to produce a single, composite image that was used for 
photometric analysis.
Figure~\ref{fig:wfpc2} shows example composite images for the
F218W, F255W, F300W, and F450W filters.
The ratio of the signals from the F450W and F606W images taken on
Dec 15 was used to normalize the F450W photometry to the same absolute
scale as the photometry taken on Nov 30.
Figure~\ref{fig:f606w} shows all the individual images (not composites) for the 
short (0.11~s) exposures taken with the F606W filter.
The electronic gain for the 0.11~s images taken through F606W filter was 
half the value used for the other WFPC2 images
because Lutetia is bright \mbox{($V \approx 10.1$)}, and we needed
the extra dynamic range provided by the lower gain setting
(0.11~s is the shortest available exposure time for the WFPC2).

\begin{figure*}
\begin{minipage}[ht]{.5\textwidth}
	\begin{center}
	\includegraphics[angle = 0,width = \textwidth]
	{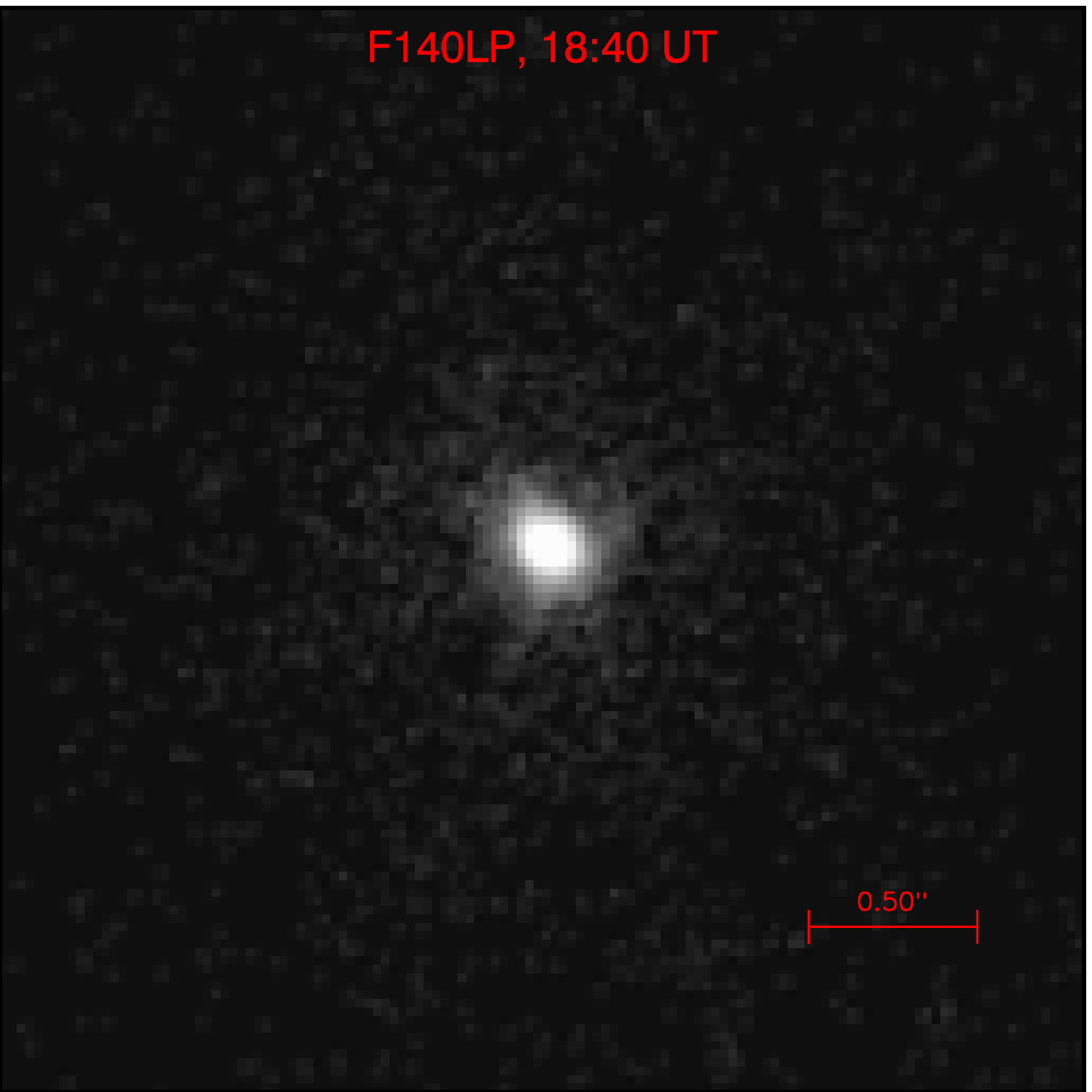}
	\end{center}
\end{minipage}
\hfill
\begin{minipage}[ht]{.5\textwidth}
	\begin{center}
	\includegraphics[angle = 0,width = \textwidth]
	{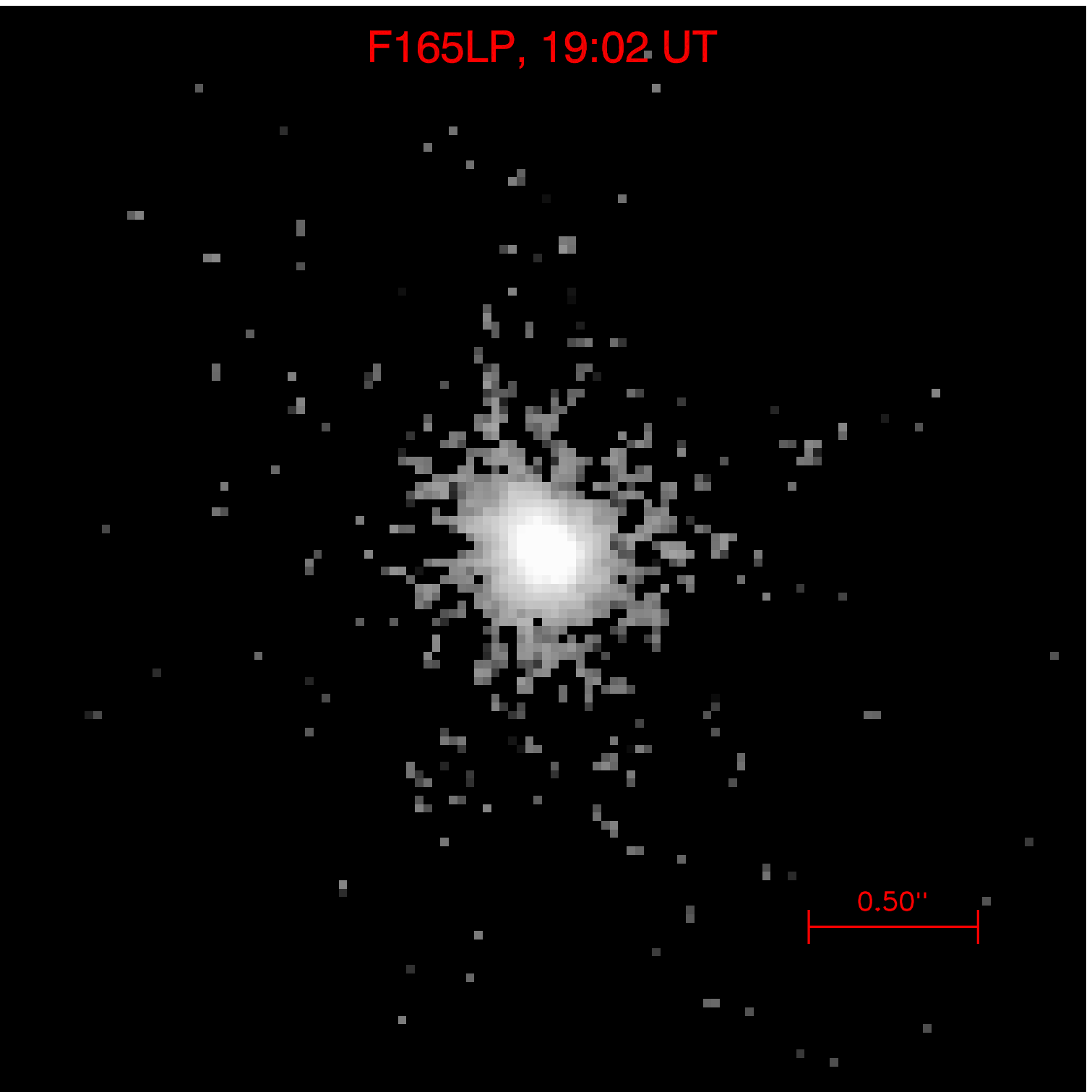}
	\end{center}
\end{minipage}
\begin{minipage}[ht]{.5\textwidth}
	\begin{center}
	\includegraphics[angle = 0,width = \textwidth]
	{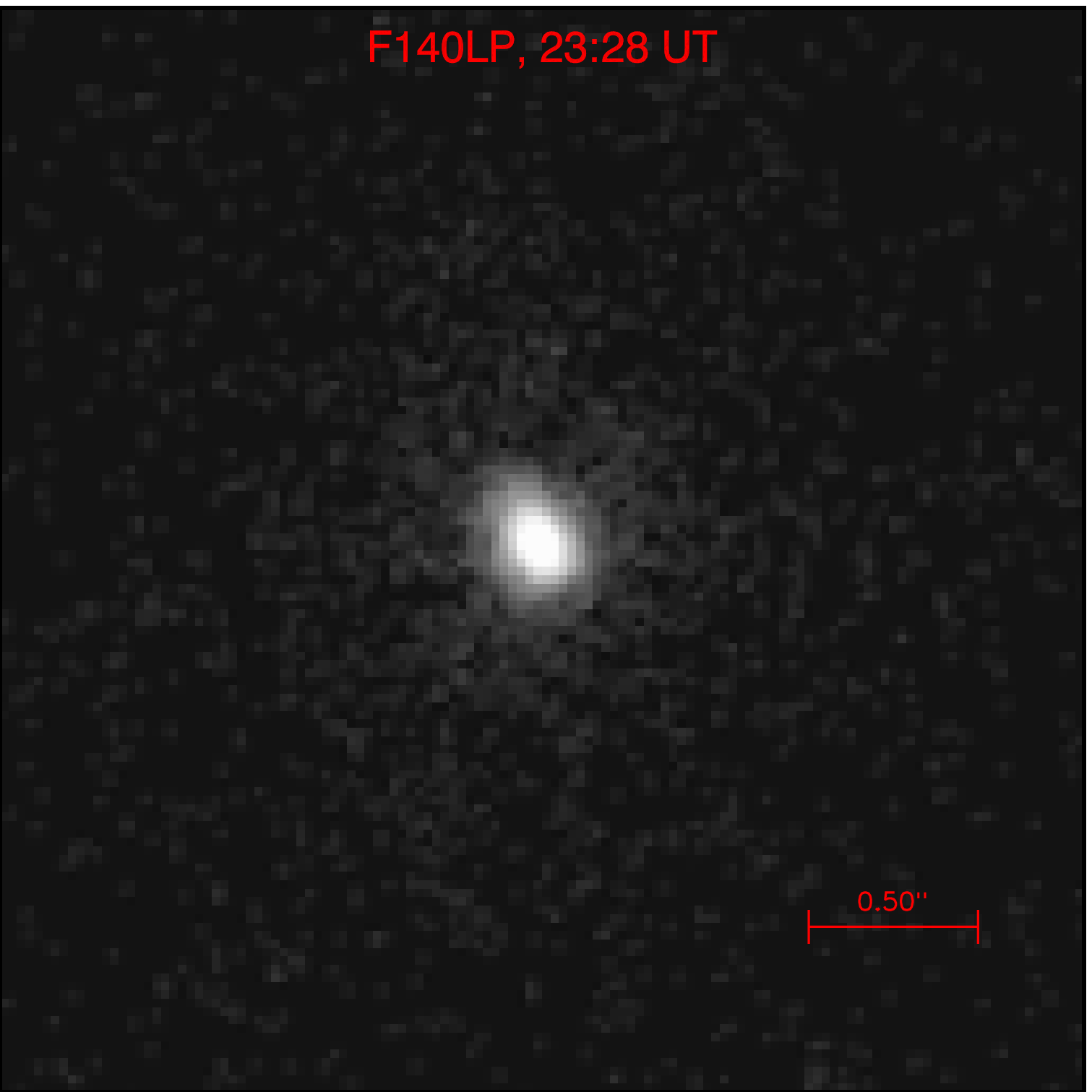}
	\end{center}
\end{minipage}
\hfill
\begin{minipage}[ht]{.5\textwidth}
	\begin{center}
	\includegraphics[angle = 0,width = \textwidth]
	{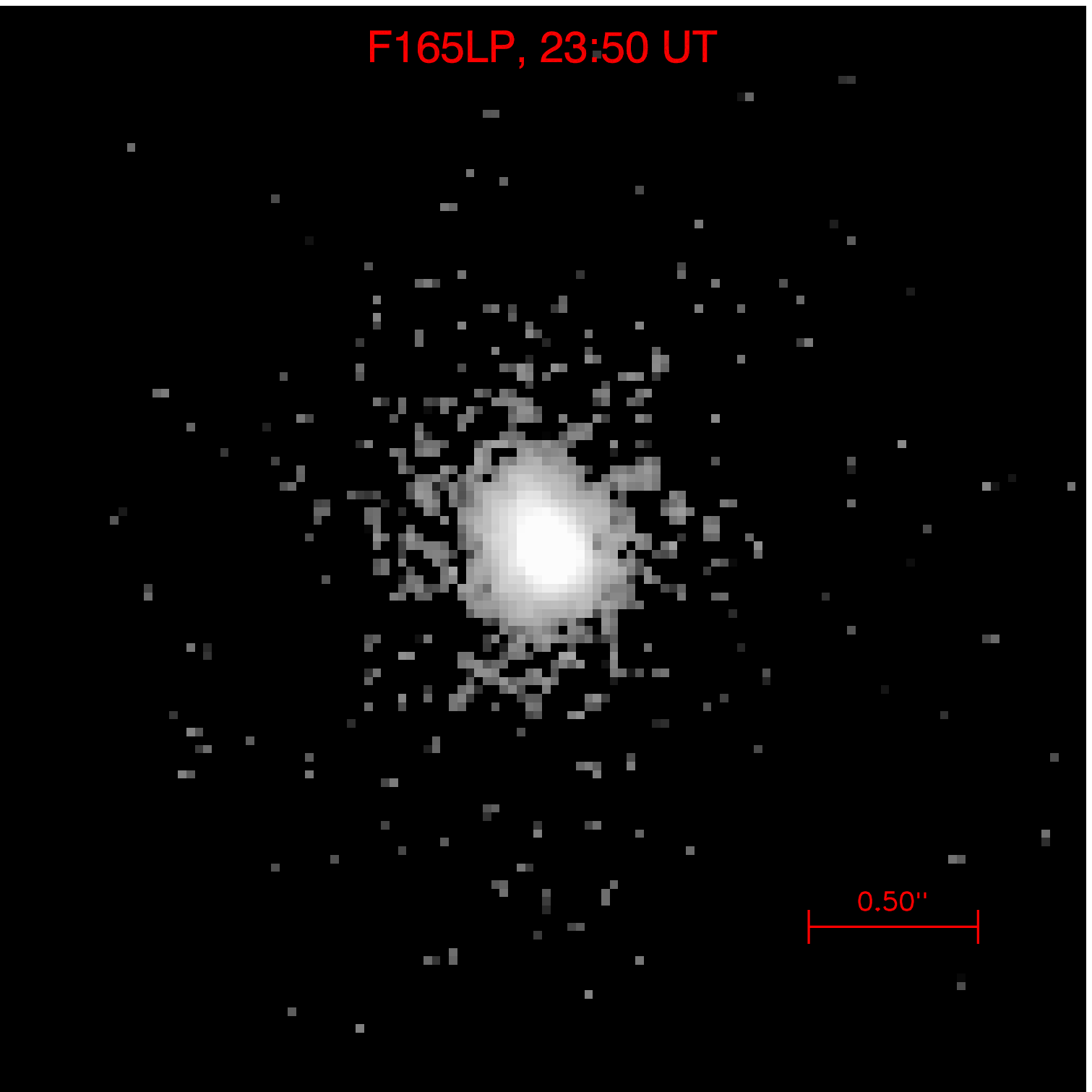}
	\end{center}
\end{minipage}
\caption{
{\it HST/ACS} images of asteroid 21 Lutetia taken through the F140LP
and F165LP filters (see identifying labels). 
Each image is displayed using an asinh
intensity stretch (similar to logarithmic) ranging approximately from zero to the 
maximum intensity in each image.
All images have been rotated so that celestial north points up and east is to the left.
The scale bar is 0\farcs5 across, which subtends $\sim$520~km on 30~November
2009 when the images were taken (start times are labeled on each image).
}
\label{fig:sbc}
\end{figure*}

The ACS/SBC employs a photon-counting detector that is essentially insensitive to cosmic
ray contamination, so we simply took a single, long exposure at a single location on
the detector for our FUV imaging.
The four images (2 for F140LP and 2 for F165LP) are displayed in
figure~\ref{fig:sbc}.
Unfortunately, a star passing near Lutetia contaminated the F165LP image taken
during visit~05, rendering the photometry unusable for that observation.
Although the FUV photometry was obtained $\sim$75~min after the photometry
taken through the other filters, we did not attempt to correct the FUV photometry for
any light curve variation because that effect was determined to be $\sim$4\%
(using the shape model from \citealt{Drummond:2010}), whereas
the statistical error in the \mbox{F140LP--F165LP} photometry is $\sim$17\% (see below).

We used standard circular aperture photometry to determine the total signal in 
each {\it HST} filter.
We then compared the observed signal  to the predicted signal using a model Lutetia albedo spectrum and measured {\it HST} throughput curves (a technique called
\emph{synthetic photometry}).
We iteratively adjusted the albedo model until the {\it measured} count rates 
from the {\it HST} filter photometry matched the {\it predicted} values
from synthetic photometry to within the measurement uncertainties.
In addition to the Hubble photometry, we used a ground-based spectrum taken on 
28~November 2008 \citep{Perna:2010},
just two days prior to the first {\it HST} observations, to constrain the
slope of Lutetia's albedo at visible wavelengths.
All albedos discussed here refer to the \emph{geometric albedo}, which
is the albedo at a solar phase angle of 0\degr.
We adopted a phase correction factor of 0.91 when converting fluxes
from a phase angle of 0\degr\ to the observed phase angle
of 0\fdg48 on 2008 November~30, which is consistent with the
phase law deduced from visible observations \citep{Belskaya:2010}.

In coordination with our {\it HST} effort, we acquired high-angular resolution
adaptive-optics imaging at the Keck Telescope
near the times of the 2008 opposition.
We combined these data with data previously acquired at the 
Very Large Telescope (VLT) near the prior opposition, in June 2007,
to  provide improved  estimates of the size, shape, and spin axis of Lutetia, 
and also to search for satellites
 \citep{Carry:2010, Drummond:2010, Merline:2010}.
We adopt here the hybrid shape model
derived by combining the results from the direct size measurements
and the inversion of various light curve data \citep{Carry:2010, Drummond:2010}.
This model is then used to predict Lutetia's projected area at the
time of the {\it HST} observations.
The predicted projected area varies between 9615~km$^{2}$ and 10081~km$^{2}$
over the course of the {\it HST} observations, and this variation of $\sim$4.8\% is within the expected uncertainty in the prediction itself ($\sim$6\%).
We have adopted a projected area of 10000~km$^{2}$, corresponding to an 
effective diameter for Lutetia of 113~km, 
to set the absolute scale for the albedo.
(Note that the \emph{mean} 3-dimensional diameter of Lutetia is 105~km, 
as determined by \citealt{Drummond:2010}. But when viewed from high latitudes, 
or close to pole-on, the effective diameter of the projected disk is larger, 
due to viewing primarily the a,b dimensions.
An independent size estimate can be derived from the {\it HST} images because Lutetia is slightly resolved, but detailed modeling is required to derive accurate size and shape
estimates, and we are deferring this work to a future paper.)
We estimate that the uncertainty in the albedo derived below is $\sim$6\%.

\begin{figure*}[ht]
\resizebox{\hsize}{!}{
\includegraphics{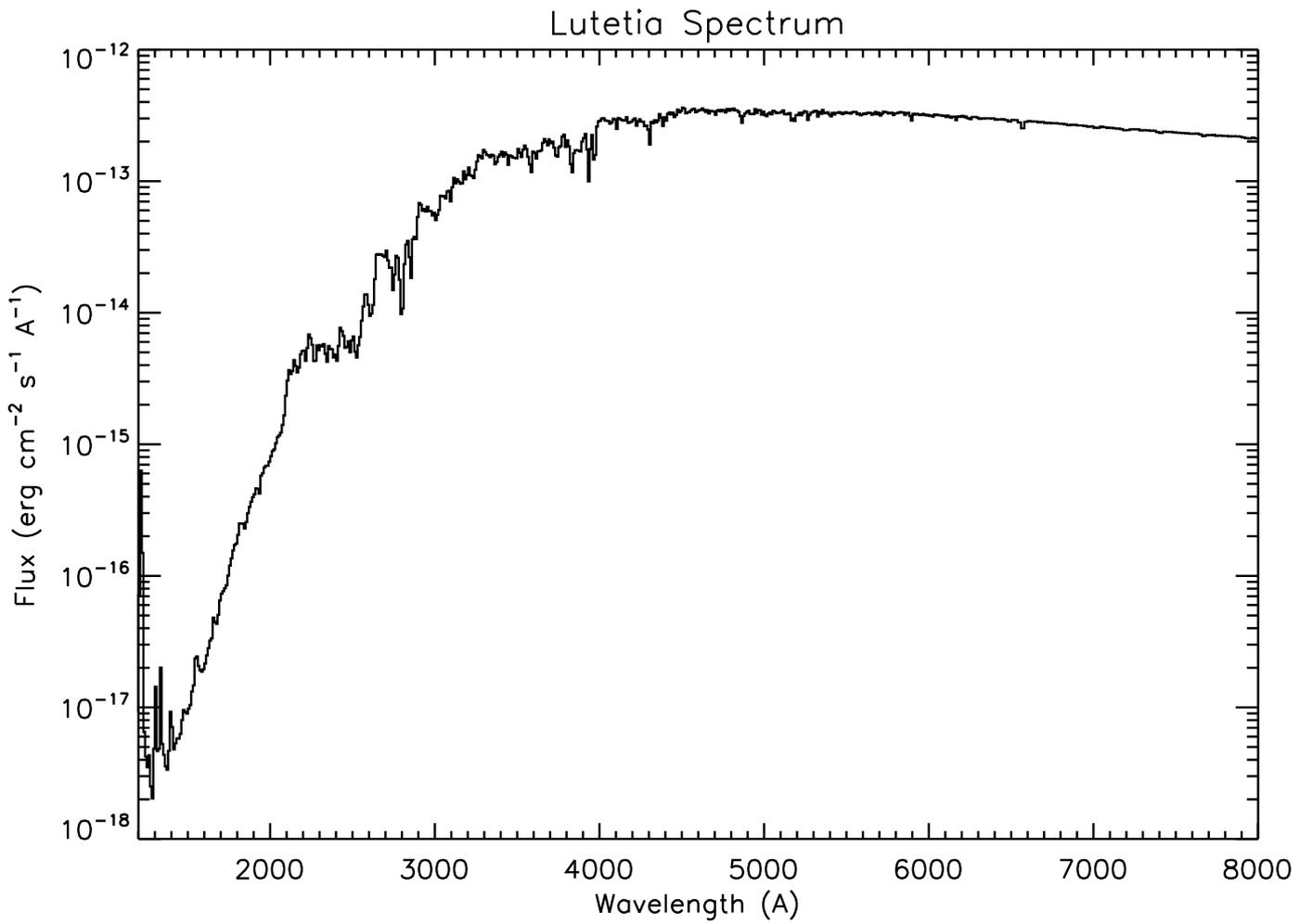}
}
\caption{Our best estimate for the Lutetia spectrum at the time of the {\it HST} observations
on 30~November 2008.}
\label{fig:spectrum}
\end{figure*}

We used the solar spectrum from \cite{Colina:1996} to convert from albedo to
absolute fluxes.
Table~\ref{table:results} summarizes the numerical results after the final iteration, and
figure~\ref{fig:spectrum} displays our best estimate for Lutetia's absolute flux
at the time of the first {\it HST} observations on 30.726~November 2008.

\begin{figure*}[ht]
\resizebox{\hsize}{!}{
\includegraphics{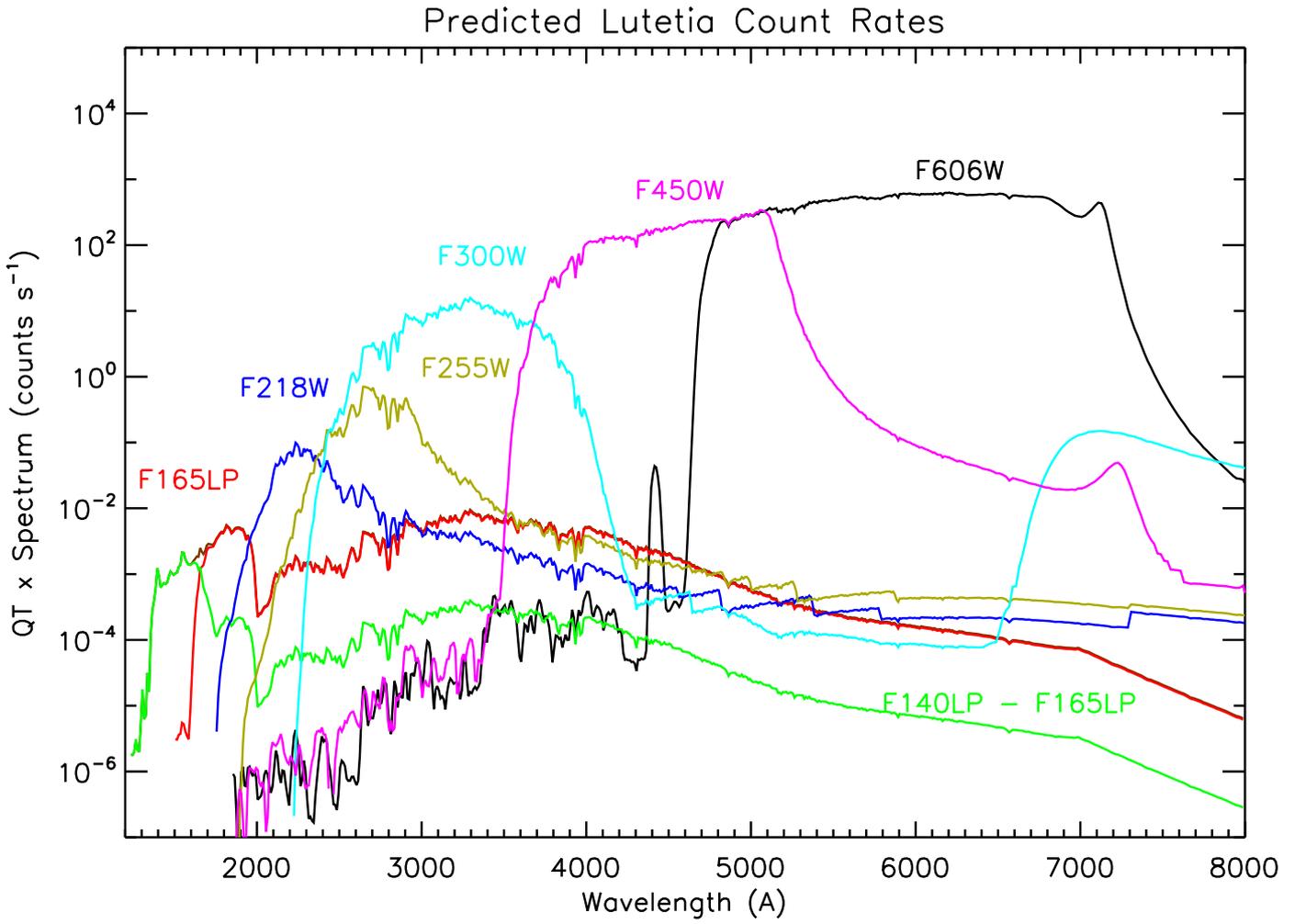}
}
\caption{After adopting our best estimate for Lutetia's flux, we plot the {\it predicted} count rates 
as a function of wavelength for each of the filters employed during the {\it HST} observations.
\mbox{``F140LP--F165LP''} refers to the {\it difference} between the F140LP and F165LP
filters.
For clarity, the F140LP curve is not explicitly labeled, but it is essentially identical
to the F165LP curve longward of 1650~\AA\ and essentially identical to the 
\mbox{F140LP--F165LP} curve shortward of that wavelength.
Note the logarithmic scale.}
\label{fig:throughputs}
\end{figure*}

\begin{figure*}[ht]
\resizebox{\hsize}{!}{
\includegraphics{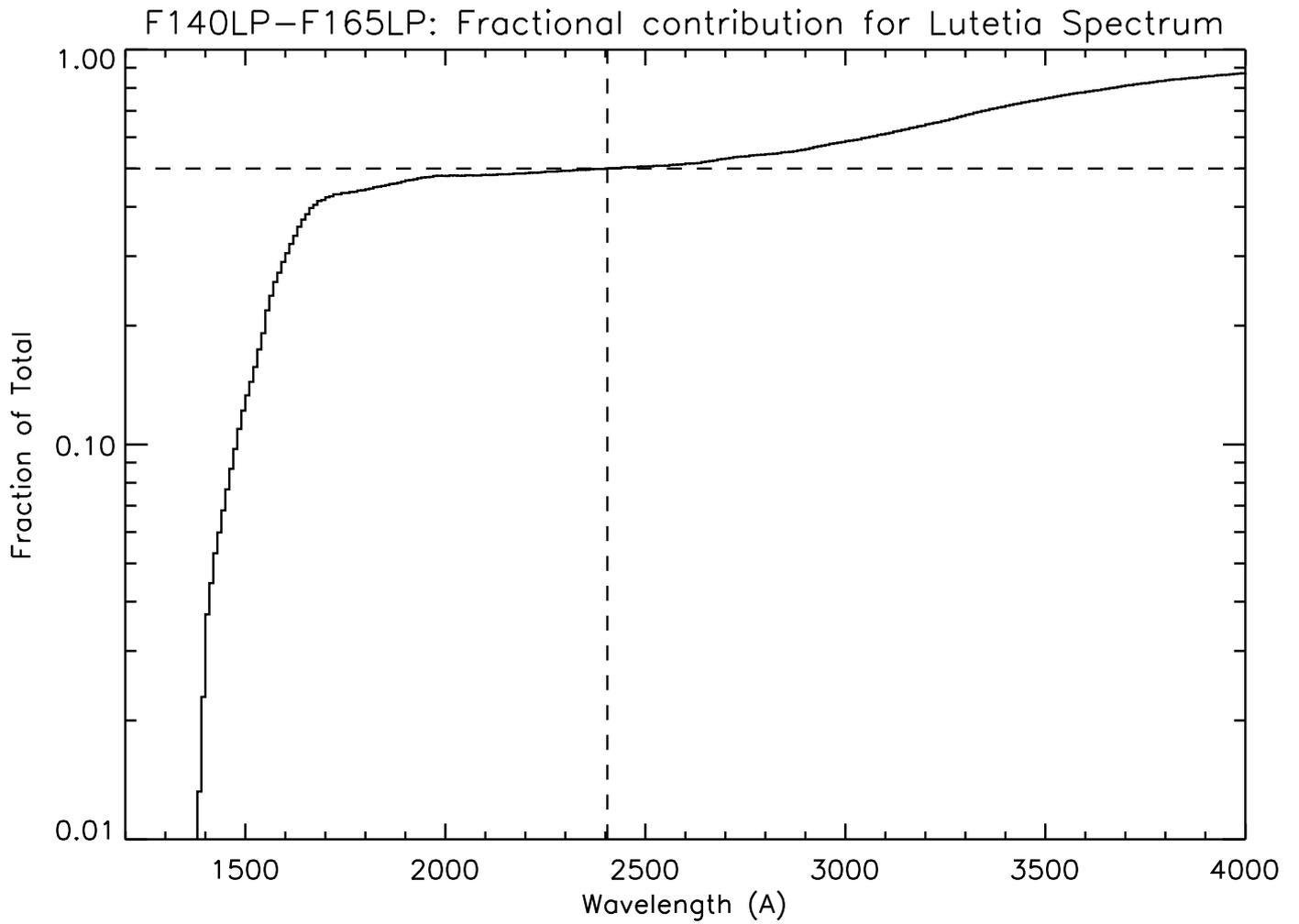}
}
\caption{The cumulative fractional contribution to the observed count rate as a function
of wavelength for the {\it difference} (F140LP--F165LP) of the two far-UV filters employed 
during the {\it HST} observations.
Approximately 40\% of the observed signal is produced by photons having wavelengths
smaller than 1675~\AA, but 50\% of the signal is coming from light longward of 2400~\AA.}
\label{fig:fraction}
\end{figure*}

Applying synthetic photometry to the {\it HST} data is essential for obtaining
accurate results because Lutetia's flux varies dramatically across the 
passbands of all the filters used, except F450W and F606W.
Figure~\ref{fig:throughputs} shows the predicted count rates as a function of
wavelength for each of the filters employed during the {\it HST} observations,
as well as the response for the difference \mbox{F140LP--F165LP.}
Determination of the FUV flux is especially problematic owing to 
``red leak'' issues (i.e., when much, or even most, of the observed signal is 
produced by photons whose wavelengths are much longer than the wavelength of the 
peak in the filter transmittance),
which figure~\ref{fig:throughputs} graphically illustrates.
For the F140LP filter, only $\sim$10\% of the observed signal is produced by photons having wavelengths shortward of 1895~\AA.
For F165LP, only $\sim$10\% of the observed signal is produced by photons having wavelengths shortward of 1975~\AA.
The 50\% point (i.e., half the observed signal is produced
by photons either shortward or longward of that wavelength) occurs
at 3340~\AA\ for F140LP and 3390~\AA\ for F160LP.
We mitigate the red leak problem by forming the difference signal (F140LP--F165LP),
which significantly improves the situation, as indicated in figure~\ref{fig:fraction}.
In that case, $\sim$40\% of the difference signal is produced by photons having
wavelengths shortward of 1675~\AA, and the 50\% point occurs at 2400~\AA.
Nevertheless, the red leak remains a major issue affecting the accuracy
of our FUV results.

In order to match the measured signals for the F140LP and
the F165LP filters, and additionally match their \emph{difference} signal 
\mbox{(F140LP-F165LP),} we had to increase
the system throughput (QT) for each of those filters by a factor of $2.5$ for 
wavelengths longward of 2000~\AA, relative to the curves currently adopted
by the STScI (i.e., the red leak is even worse than originally thought).
After consultation with the relevant experts at the STScI, we concluded
that the large uncertainty in the long wavelength response of the FUV
filters justifies our empirical approach.
While our choice for the filter throughputs is certainly not unique,
and may be incorrect in detail, we are confident in our assessment
that a red leak adjustment for the F140LP and F165LP filters,
of approximately the magnitude adopted here, is required
to produce consistent results.

\begin{figure*}[ht]
\resizebox{\hsize}{!}{
\includegraphics{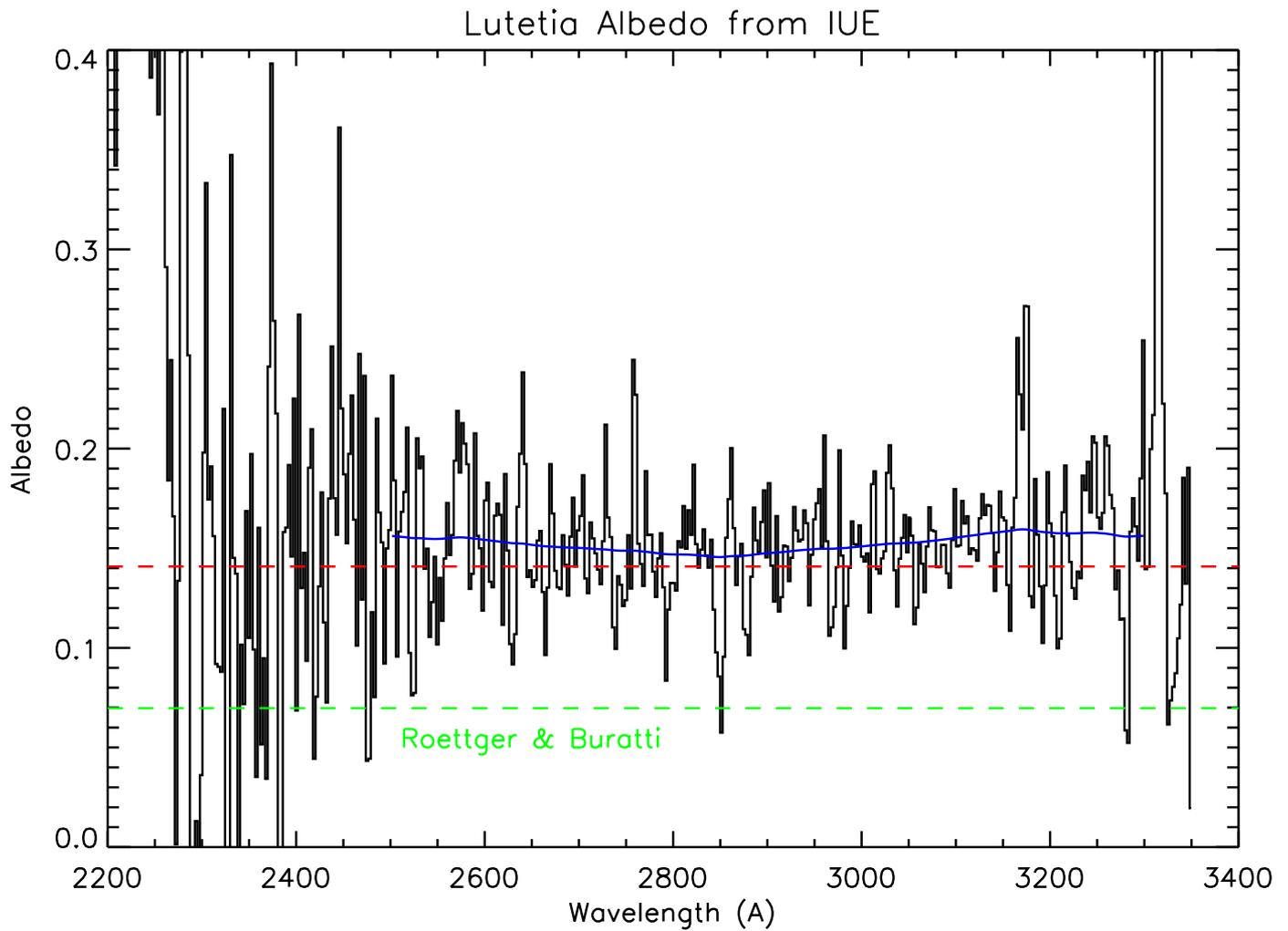}
}
\caption{Albedo spectrum of Lutetia derived from {\it IUE} observations made on
7~January 1982 at a solar phase angle of 26\fdg1. The albedo at 2670~\AA\ 
adopted by \cite{Roettger:1994} is shown by the dashed green line, while our 
equivalent value, using a different phase correction, is depicted by the dashed red line.
The blue curve is a Fourier-filtered version of the {\it IUE} spectrum, passing only
the lowest 1\% of spatial frequencies.}
\label{fig:iue}
\end{figure*}

Additional information on Lutetia's UV albedo is available from {\it IUE} observations
performed in 1982, originally analyzed by \cite{Roettger:1994}.
We obtained these data from the STScI archive, and reanalyzed them using
new information on the phase behavior \citep{Belskaya:2010} and size of Lutetia
\citep{Drummond:2010}.
The {\it IUE} spectrum (figure~\ref{fig:iue}) is noisy and was taken at a solar phase 
angle of 26\fdg1, making the correction to geometric albedo rather uncertain.
Adopting a phase correction factor of 3.1, which is the value observed for Lutetia 
at visible wavelengths, we obtain an average albedo of $\sim$0.14 near 2670~\AA.
The latter is double the value adopted by \cite{Roettger:1994} (0.074, after correction 
to the new effective diameter), apparently
owing to their use of a different phase law.
The phase law typically depends on both the absolute albedo and the wavelength.
Thus, it is not surprising that our phase correction factor is significantly different than
the one adopted by  \cite{Roettger:1994}, especially since there is scant data
available on the phase behavior of Lutetia at UV wavelengths.
In order to match the {\it HST} photometry from the F255W filter, we adopted
an albedo of $\sim$0.10 near 2670~\AA, which is approximately halfway between
the two different {\it IUE} results.

Of greater concern is the slope of Lutetia's albedo between 2400~\AA\ and
3300~\AA.
The {\it HST} data suggest there is a sharp drop in Lutetia's albedo in the wavelength
range $\sim$3000-3300~\AA, while the {\it IUE} data indicate that the albedo is essentially
constant over the wavelength range $\sim$2400-3300~\AA.
Although the {\it IUE} observations were performed nearly 27 years before the
{\it HST} observations, we would not expect the UV albedo of Lutetia to change either
in its slope or its absolute value over that time.
The aspect angles of the {\it HST} and {\it IUE} observations were significantly
different (Lutetia's sub-Earth latitude was $-$73\degr during the {\it HST} observations
and $-$51\degr during the {\it IUE} observations), and perhaps albedo variation
over Lutetia's albedo could explain the differences in the {\it HST} and {\it IUE} results.
However, the long exposure time for the {\it IUE} spectrum (3.2~hr) covered nearly
75\% of Lutetia's lightcurve period, suggesting that surface variegation may not 
play a significant role.
In summary, there appears to be a discrepancy between the
{\it IUE} and {\it HST} results for the slope of Lutetia's UV albedo near
3100~\AA.
Nevertheless, both the {\it IUE} and {\it HST} data indicate that the NUV albedo
is significantly smaller than the visible albedo.
The  {\it HST} data further suggest that the FUV albedo is approximately
60\% of the visible albedo.

\section{Discussion}

Lutetia was extensively observed in the 1970s, yielding visible and
near-IR reflectance spectra \citep{McCord:1975a},
radiometric albedos and diameter estimates \citep{Morrison:1977a}, 
and polarimetric albedos and diameter estimates \citep{Zellner:1976a},
which have been confirmed by similar observations reported during the last decade
(see \citealt{Belskaya:2010}). Based on these data,
\cite{Chapman:1975} assigned Psyche, Lutetia, and
Kalliope to a new, distinct taxonomic group, to which
\cite{Zellner:1976a} assigned the letter ``M''. 

The M-type was defined in terms of spectral and albedo properties
by \cite{Bowell:1978}.
Although the visual albedos for M-class objects covered a
broad range ($\sim$7--23\%), they were larger
than the albedos for typical carbonaceous chondrites (CC)
and overlapped the range reported for S-types.
Recent radiometry of Lutetia, reduced using two different
thermal models, confirmed that Lutetia's visual albedo 
is higher than for typical CC meteorites \citep{Mueller:2006}.
However, some types of CCs (CO/CV) have visual albedos
approaching $\sim$16\% \citep{Clark:2009}, essentially
identical to the value derived here for Lutetia from
the {\it HST} observations.

\cite{Rivkin:1995} recognized that there were two sub-classes
of M-type asteroids.
The standard M types had high radar albedos and relatively
neutral visible colors, properties that could be attributed to 
metal-dominated surfaces.
But some of the M type asteroids, including Lutetia, had an absorption band
near 3~$\mu$m band, which was attributed to
hydrated minerals.
\cite{Rivkin:2000} call this new ``wet'' class M(W) and assigned
Lutetia to it.

Radar observations of Lutetia
\citep{Magri:1999, Magri:2007}, confirmed by \cite{Shepard:2008a},
showed that Lutetia has a moderate radar albedo: nominally $\sim$20\%,
but possibly as low as 10\% or as high as 31\% after accounting
for the error bars. Lutetia's radar albedo is lower than for the largely metallic
asteroids but is somewhat higher than for typical CCs
(13\% $\pm$ 5\% for the average C-class object, according to \citealt{Shepard:2008a}).

Numerous researchers in the last few years (\citealt{Barucci:2005, Barucci:2008,
Lazzarin:2009, Perna:2010}; see summary by \citealt{Belskaya:2010})
have argued that Lutetia shows certain spectral
characteristics ({\it e.g.,} in the thermal IR) that resemble several
CO and CV meteorites, but not an iron meteorite.
However, mineralogical interpretations of thermal IR spectra must
be made cautiously because particle size, in addition to composition,
can strongly affect the observed spectral features \citep{Vernazza:2010}.
We further note that: (a)  the lack of a drop-off
in Lutetia's spectral reflectance below 0.55~$\mu$m and its relatively
high albedo make it inconsistent with CV meteorites 
\citep[see][for instance]{Gaffey:1976}, and
(b)  CO meteorites display a 1~$\mu$m
olivine band that is absent in Lutetia's reflectance
spectrum \citep[see Fig. 3 of][]{Barucci:2005}.
 
It was first suggested by \cite{Chapman:1973} that what we now term an
M-type spectrum might be associated with enstatite chondrites (ECs).
More recently, \cite{Rivkin:2000} have suggested that a hydrated
EC is a plausible composition for Lutetia, consistent with
the recent analysis of \citealt{Vernazza:2009b}.
From rotationally resolved visible and near-IR spectra of Lutetia,
\cite{Nedelcu:2007a} claimed a better match with CC in one hemisphere
and with EC in the other, but this hemispherical spectral asymmetry
has not yet been confirmed by other researchers.

Recent dynamical work \citep{Baer:2008, Fienga:2009} has provided 
an estimate for Lutetia's mass,
which combined with the new size estimates \citep{Drummond:2010, Carry:2010}
yield a bulk density of \mbox{$\sim$4 g cm$^{-3}$}
(the formal uncertainty ranges from \mbox{2.4--5.1 g cm$^{-3}$};
see \citealt{Drummond:2010}).
This density is too small for an object having a dominantly metal component
and seems more compatible with an EC-like composition \citep{Drummond:2010}.

As discussed by \cite{Roettger:1994}, the slope of Lutetia's NUV albedo is similar
to that of the M- and S-type asteroids observed by {\it IUE}. The albedo of the C-type asteroids
observed by {\it IUE} increases (by $\sim$10\%) between 2400~\AA\ and 3000~\AA, whereas
there is little to no variation in Lutetia's albedo over this wavelength range.
Both the {\it IUE} and {\it HST} data demonstrate that the absolute value
of Lutetia's NUV albedo is larger than typically observed for the C-type asteroids.

According to the {\it HST} data, Lutetia has a rather high FUV albedo of $\sim$10\%
over the entire wavelength range from $\sim$1500~\AA\ to $\sim$3000~\AA.
This can be compared to an FUV albedo of $\sim$4\% for the Earth's Moon
\citep{Henry:hutmoon} and the E-type asteroid 2867~Steins \citep{Ahearn:steins}, 
the only asteroid observed at FUV wavelengths. Furthermore, neither the Moon
nor Steins show an abrupt drop in albedo near 3200~\AA.

We compared Lutetia's albedo to laboratory reflectivity measurements 
of a wide variety of materials, including meteorite and lunar samples \citep{Wagner:1987},
and none of them appear to be good analogs for Lutetia's surface.
The feldspar powders have a sharp albedo drop near 3000~\AA\ and
have NUV-FUV albedos similar to that of Lutetia, but the ratios between their visible 
and UV albedos are several times larger than we find for Lutetia.
SO$_{\rm 2}$ frost also has a sharp drop in albedo near 3000~\AA\ and
an FUV albedo in the range of 10-15\%, both of which are consistent with
Lutetia's UV spectrum. However, the visible-to-UV albedo ratio of SO$_{\rm 2}$ frost 
is several times larger than Lutetia's, and exposed frost isn't expected 
to be present on Lutetia's surface.
The albedos of chondritic meteorite samples are significantly smaller than
Lutetia's albedo at all wavelengths, in addition to not matching Lutetia's
spectral variations.
Similarly, lunar samples tend to have lower UV albedos than Lutetia.
Spectra of various mineral powders (e.g., iron, clays, sulfur) also show
striking differences when compared to Lutetia's spectrum.
Perhaps some mixture of samples could be found to approximate
Lutetia's spectrum, but such an effort is beyond the scope of this paper.
We note, however, that many of the materials measured by
\citet{Wagner:1987} have spectral features shortward of 2000~\AA, 
which are potentially observable by
the {\it Alice} instrument during the {\it Rosetta} flyby.

\section{Conclusion}

Using the \emph{Hubble Space Telescope}, we measured the albedo of 
asteroid (21) Lutetia over a wide wavelength range, extending from
the far ultraviolet ($\sim$1500~\AA) to the visible ($\sim$6000~\AA).
The {\it HST} results reported here suggest a sharp drop in Lutetia's albedo near
3100~\AA, and an essentially constant FUV albedo of $\sim$10\% between
1400--3000~\AA.
The absolute value and spectral variation of Lutetia's UV-visible albedo is not well-matched
by the spectra of any meteorites measured in the laboratory.
Lutetia may well be composed of material that is
either rare or not yet represented in our meteorite collections.

Lutetia's FUV albedo is considerably higher than the values measured 
for C-chondrites and the Earth's Moon ($\sim$4\%), which implies that
Lutetia should be a relatively easy target for the {\it Alice} instrument when
it makes observations during the {\it Rosetta} close flyby in July~2010.

\begin{acknowledgements}
We thank the STScI Director, Matt Mountain, for granting us discretionary
observing time for the {\it HST} Lutetia observations.
We thank the STScI ground system personnel, and especially Alison Vick, for successfully
planning and executing the observations.
We gratefully acknowledge Jennifer Mack and John Biretta for discussions on
``red leaks'' and for providing the best available {\it ACS/SBC} and {\it WFPC2} system
throughputs.
We thank the other members of the Merline-Drummond
ground-based observing team, particularly Al Conrad
and Beno\^{i}t Carry, for providing results on the size
and shape of Lutetia prior to publication.
The comments of an anonymous reviewer were helpful and have led to 
several improvements in the paper.
\end{acknowledgements}

\bibliographystyle{aa} 	
\bibliography{bibtex_refs} 	

\clearpage

\begin{table*}[ht]
\caption{Log of {\it HST} observations of Lutetia.} 
\label{table:log} 
\centering 
\begin{tabular}{l l l}
\hline\hline 
Visit Info & Measurements & Objectives \\ 
\hline 
ub9h01xxx & WFPC2/PC, 2 dither points & Near-UV albedo \\ 
2008-Nov-30 & F218W, 4 $\times$ 160 s & Bridge from FUV to Visible albedo \\
17:05-17:45 UT & F255W, 4 $\times$ 40 s & \\
 	& F300W, 2 $\times$ 2 s & \\
	& F606W, 2 $\times$ 0.11 s & \\
\hline
ub9h02xxx & ACS/SBC & Far-UV albedo for one hemisphere \\
2008-Nov-30 & F140LP, 1 $\times$ 1270 s & \\
18:40-19:25 UT & F165LP, 1 $\times$ 1270 s & \\
\hline
ub9h03xxx & WFPC2/PC, 2 dither points & Visible albedo \\
2008-Nov-30 & F606W, 2 $\times$ 0.11 s & Deep probe for dust debris \\
20:17-20:58 UT & F606W, 3 $\times$ 40 s & Deep probe for companions \\
 	& F606W, 7 $\times$ 160 s & \\
\hline
ub9h04xxx & WFPC2/PC, 2 dither points & Verify putative companions \\
2008-Nov-30 & F606W, 2 $\times$ 0.11 s & Repeat of ub9h03xxx\\
21:52-22:33 UT & F606W, 3 $\times$ 40 s &  \\
 	& F606W, 7 $\times$ 160 s & \\
\hline
ub9h05xxx & ACS/SBC & Far-UV albedo for opposite hemisphere \\
2008-Nov-30 & F140LP, 1 $\times$ 1270 s & \\
23:27-00:13 UT & F165LP, 1 $\times$ 1270 s & \\
\hline
ub9h13xxx & WFPC2/PC, 2 dither points & \emph{B-V} color of Lutetia and companions \\
2008-Dec-15 & F606W, 2 $\times$ 0.11 s, 2 $\times$ 40 s, 2 $\times$ 160 s & \\
13:26-14:08 UT & F450W, 2 $\times$ 0.35 s, 4 $\times$ 140 s &  \\
\hline
ub9h14xxx & WFPC2/PC, 2 dither points & Verify colors \\
2008-Dec-16 & F606W, 2 $\times$ 0.11 s, 2 $\times$ 40 s, 2 $\times$ 160 s & \\
13:26-14:06 UT & F450W, 2 $\times$ 0.35 s, 4 $\times$ 140 s &  \\
\hline 
\end{tabular}
\end{table*}

\clearpage

\begin{table*}[ht]
\caption{{\it HST} photometry of Lutetia.} 
\label{table:results} 
\centering 
\begin{tabular}{c c c c}
\hline\hline 
Filter & \multicolumn{2}{c}{Measured Signal (e s$^{-1}$)} & Model Signal \\ 
 & Total & Error & (Total e s$^{-1}$) \\
\hline 
 F140LP & 12.52 & 0.10 & 12.52 \\
 F165LP & 11.69 & 0.10 & 11.64 \\
 F140LP-F165LP & 0.828 & 0.141 & 0.879 \\
 F218W & 33.7 & 1.7 & 33.4 \\
 F255W & 226 & 7.0 & 226 \\
 F300W & 9625 & 150 & 9499 \\
 F450W & 120,160 & 2400 & 122,990 \\
 F606W & 1.107 $\times$ 10$^{6}$ & 4.42 $\times$ 10$^{3}$ & 1.122 $\times$ 10$^{6}$ \\
 \hline
 \end{tabular}
\end{table*}

\clearpage

\clearpage

\clearpage

\clearpage

\clearpage

\clearpage

\clearpage

\clearpage

\clearpage

\begin{figure*}[ht]
\resizebox{\hsize}{!}{
\includegraphics{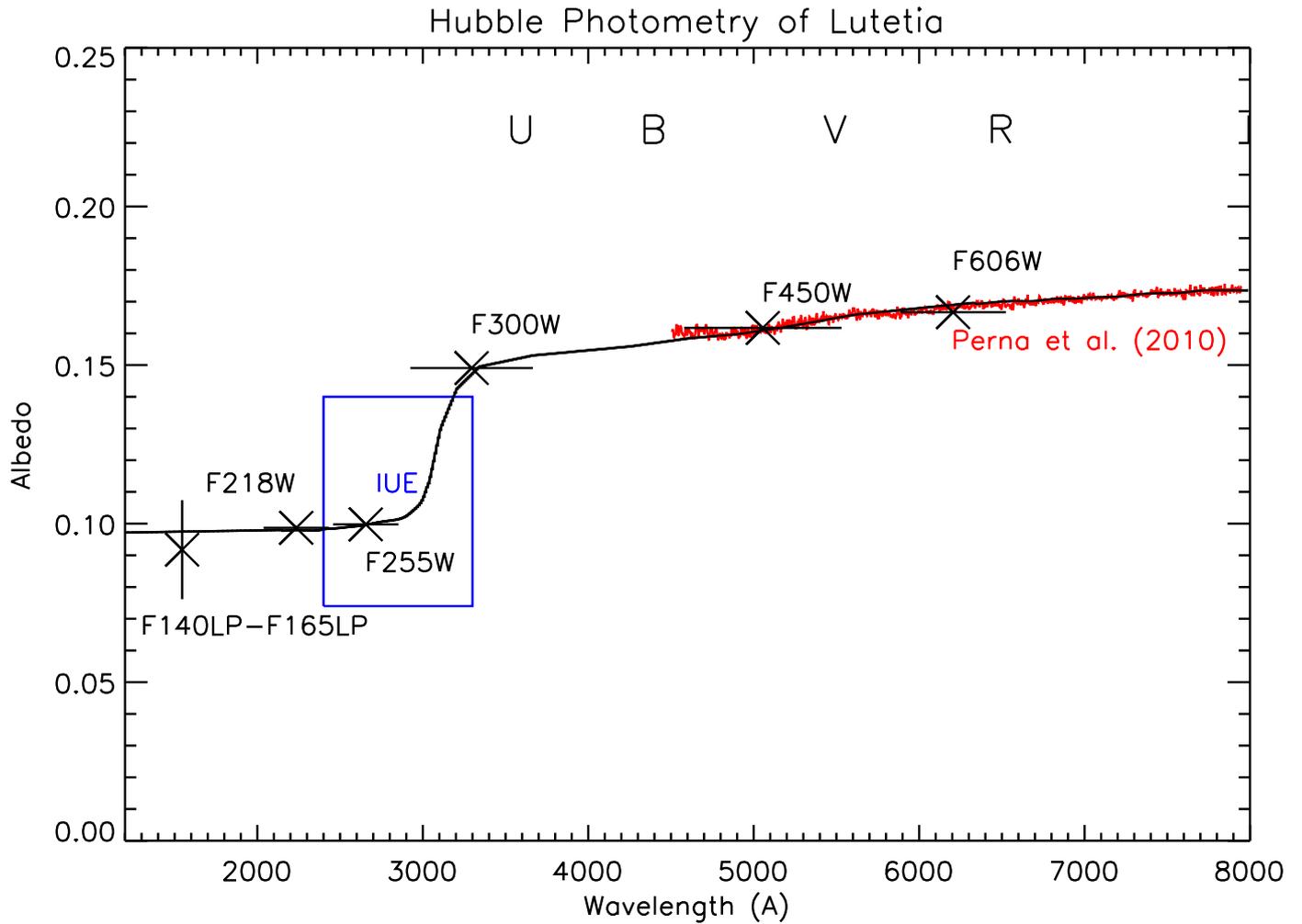}
}
\caption{Our best estimate for the albedo of Lutetia is plotted as a solid black line.
The spectrum plotted in red is from a ground-based observation taken on
28~November 2008.
The blue rectangle shows the wavelength coverage and the range of possible albedos 
derived from an {\it IUE} spectrum taken on 7~January 1982.
The {\it HST} data ($\times$) are plotted at the wavelengths where the predicted
count rate is largest; the horizontal bars give the ``effective bandwidth'' of the filters, 
as specified in the STScI Instrument Handbooks.
The wavelengths of the standard {\it UBVR} bands are also displayed for reference.
Only the error bar for the \mbox{F140LP--F165LP} {\it difference} filter case is 
displayed because the error bars for the other filters are smaller than the plotting symbols.
}
\label{fig:albedo}
\end{figure*}

\end{document}